\documentclass[10pt,letterpaper]{article}
\usepackage{graphicx}
\usepackage{amsmath}
\usepackage{amsfonts}
\usepackage{subcaption}
\usepackage{multirow}
\usepackage{booktabs} 
\usepackage{xcolor}
\usepackage[numbers,sort&compress]{natbib}
\usepackage[hidelinks]{hyperref}
\usepackage{amsmath}
\usepackage{xspace}
\usepackage{tabularx}
\usepackage{bm}
\usepackage[font={footnotesize}]{caption}
\usepackage{authblk}

\usepackage[switch]{lineno}

\hypersetup{colorlinks,linkcolor={magenta},citecolor={blue},urlcolor={magenta}}

\title{A temporal graph dissimilarity measure\\ capturing epidemic spreading behavior}

\author[1,2,*]{Salvatore Freni}
\author[1,*]{Lorenzo Dall'Amico}
\author[2]{Francesca Pistilli}
\author[3]{Alain Barrat}
\author[2]{Tatiana Tommasi}
\author[1]{Ciro Cattuto}

\affil[1]{\small \textit{ISI Foundation, Turin, Italy}}
\affil[2]{\small \textit{Politecnico di Torino, Turin, Italy}}
\affil[3]{\small \textit{Aix Marseille Univ, Univ Toulon, CNRS, CPT, Turing Center for Living Systems, Marseille, France}}
\affil[*]{{\small Corresponding authors: \texttt{salvatore.freni@polito.it}, \texttt{lorenzo.dallamico@isi.it}}}

\date{}

\newcommand{\name}{\emph{Epidemic-Aware Temporal Graph Dissimilarity}\xspace}
\newcommand{\ie}{\emph{i.e.}, }
\newcommand{\acronym}{EATGD\xspace}

\begin{document}

\maketitle

\begin{abstract}
Comparing temporal networks is a challenging problem in network analysis. Commonly adopted methods compare structural properties of temporal graphs and can be highly sensitive to the pre-selected features. However, temporal networks are the support of dynamical processes, and a process may take place in statistically indistinguishable ways on structurally different temporal graphs. In particular, a dissimilarity measure that compares temporal graphs based on how spreading processes unfold on them is highly desirable. In this work, we first define the \emph{Time-Shift DTW}, a dissimilarity between epidemic curves that accounts for both differences in their shape and temporal shifts. We then introduce \name (\acronym), a deterministic and differentiable dissimilarity between temporal graphs that combines an embedding-based temporal graph distance with a set of topological and temporal features. Its parameters are calibrated using stochastic SIR simulations to maximize the correlation with the \emph{Time-Shift DTW}, after which graph comparisons require no further epidemic simulations. Using empirical human proximity networks from the SocioPatterns collaboration and synthetic temporal graphs, we show that \acronym captures epidemic-based graph dissimilarity and generalizes across heterogeneous network settings. Its differentiability and computational efficiency make \acronym suitable for integration into machine-learning pipelines and provide a potential loss function for training generative models to reproduce epidemic dynamics without requiring an exact reconstruction of fine-grained network structure.

\end{abstract}

\section{Introduction}

The complex and heterogeneous structure of human proximity interactions shapes the spread of respiratory pathogens, such as SARS-CoV-2 and influenza-like illnesses \cite{mossong2008social, leung2021transmissibility, barrat2008dynamical}. This heterogeneity includes large variations across individuals in the number of interactions \cite{barrat2004architecture, cattuto2010dynamics, onnela2007structure}, as well as in their temporal evolution,  in terms of event and inter-event durations \cite{cattuto2010dynamics, barabasi2005origin}. Heterogeneity further arises from memory effects that influence the speed of epidemic spreading \cite{karsai2011small, karsai2014time, vestergaard2014memory, rosvall2014memory, guerrini2026modeling}. Understanding and modeling these properties is therefore essential to accurately describe epidemic dynamics at the individual scale. 

\medskip
	
High-resolution data on human proximity interactions, collected in a variety of contexts, are becoming increasingly available \cite{cattuto2010dynamics, salathe2010high, stopczynski2014measuring} and can be represented with temporal graphs. Nevertheless, existing datasets remain limited in number, size, and duration. Generative models for synthetic temporal graphs can be used to augment these data, or to generate data for previously unseen contexts \cite{longa2024generating, cencetti2025generating, calmon2024preserving, dymond, temporal_motif, backbones, backbones_2, andres2024detecting}.  Fitting or evaluating generative models, however, requires assessing the similarity between pairs of temporal graphs, which is an open challenge. Existing methods often rely on comparisons of high-resolution interaction patterns and may therefore not generalize to out-of-sample contexts \cite{wills2020metrics,tantardini2019comparing, sanfeliu1983distance, koutra2013deltacon, berlingerio2012netsimile, bagrow2019information, tsitsulin2018netlsd}. Here we adopt a more parsimonious approach that compares pairs of temporal graphs based on how an infectious disease propagates on them. This comparison does not depend on the full graph structure, but focuses only on the temporal evolution of the number of infected individuals, which encapsulates multiple key epidemiological measures, including the peak prevalence, peak timing, and final attack rate.  Our approach is motivated by the key observation that distinct graphs may produce indistinguishable epidemic dynamics, as shown in related works on surrogate models for temporal networks \cite{cencetti2025generating, calmon2024preserving, longa2024generating, dyane, genois2015compensating, sapienza2018estimating}. Therefore, when using proximity interactions for downstream applications in epidemic modeling, comparing temporal graphs based on the spreading dynamics they support provides a more parsimonious and generalizable method. However, this approach is also more computationally demanding by design, as it requires simulating a large number of epidemic realizations across a broad parameter space to reliably assess graph dissimilarity.
	
\medskip

To fill this gap, we propose the \name (\acronym), an efficient, deterministic and differentiable dissimilarity measure that compares temporal graphs based on their epidemic behavior. We do so in two steps. First, we introduce \emph{Time-Shift DTW}, a modified Dynamic Time Warping \cite{dtw} dissimilarity that is explicitly sensitive to temporal lags. This measure captures the full shape of the epidemic trajectory, and is sensitive to key epidemiological quantities such as the height and timing of the infection peak, and the outbreak duration \cite{hatchett2007public, epi_features}. However, using the \emph{Time-Shift DTW} to compare temporal graphs requires first generating epidemic curves through numerous stochastic SIR simulations. We thus define \acronym to bypass this computational bottleneck. It builds on a baseline temporal graph distance \cite{Dall_Amico_2024} and corrects it with feature-specific terms, whose weights are trained to maximize the correlation with the \emph{Time-Shift DTW} obtained from the SIR stochastic simulations.

Through extensive experiments on empirical human proximity data collected by the SocioPatterns collaboration \cite{cattuto2010dynamics}, and on synthetic temporal graphs, we show that \acronym accurately captures epidemic dissimilarity. Once calibrated, \acronym avoids stochastic epidemic simulations and could be used as a loss function to train generative models that produce temporal graphs exhibiting similar epidemic dynamics without necessarily replicating the fine-grained structural details of the training data.

\section{Results}

We divide our proposed framework into two main phases. First, we introduce the \emph{Time-Shift DTW}, a dissimilarity measure between pairs of time series representing the number of infected individuals. Unlike commonly adopted approaches that only compare distributions of final epidemic sizes \cite{cencetti2025generating, calmon2024preserving, longa2024generating, dyane, genois2015compensating, sapienza2018estimating}, our measure accounts for the entire shape of the infected time series. Building on this ground truth, we then define the \name (\acronym). We calibrate this deterministic dissimilarity measure between temporal graphs such that it correlates with the \emph{Time-Shift DTW} between stochastic epidemic curves generated by SIR simulations performed on each graph \cite{tillett1992infectious}.


\subsection{Time-Shift DTW epidemic dissimilarity}

One commonly adopted method of comparing sets of epidemic simulations is to evaluate the Jensen-Shannon or Kolmogorov-Smirnov divergences between the distributions of epidemic sizes (the number of individuals that contracted the disease) across multiple stochastic realizations of the infection process 
\cite{cencetti2025generating, calmon2024preserving, longa2024generating, dyane, genois2015compensating, sapienza2018estimating}. While relevant, this approach neglects some key epidemiological quantities that depend on the \emph{shape} of the curve of infected individuals, such as the peak position and height, which play a crucial role in determining the pressure on the healthcare system \cite{anderson2020will}, or the onset time and outbreak duration \cite{hatchett2007public}. These features can be captured by looking at the entire time series of the number of infected individuals rather than the attack rate alone, which might be insensitive to substantial changes in the epidemic curve.

\medskip

We introduce a measure, based on Dynamic Time Warping (DTW) \cite{dtw}, to quantify the dissimilarity between two time series representing the number of infected individuals. DTW measures the dissimilarity between two time series by finding an optimal matching path between them and is particularly well suited to our purpose because it handles time series of different durations, such as epidemic curves. Moreover, DTW compares the entire epidemic trajectories and is sensitive to key epidemiological features such as the timing, height, and width of infection peaks, as well as the presence of multiple waves.

However, DTW is, by construction, invariant under global temporal shifts: two curves differing only by a constant delay are assigned zero DTW distance, independently of the magnitude of the delay itself. In our setting, such delays are epidemiologically meaningful, since differences in the timing of epidemic peaks correspond to different spreading dynamics.  To address this limitation, we  introduce \emph{Time-Shift DTW}. Following Ref.~\cite{lead_lag}, we define a \emph{normalized lead--lag} quantity that explicitly captures the temporal offset between two warped epidemic curves. 
\begin{figure}[!t]
	\centering
	\includegraphics[width = 0.75\columnwidth]{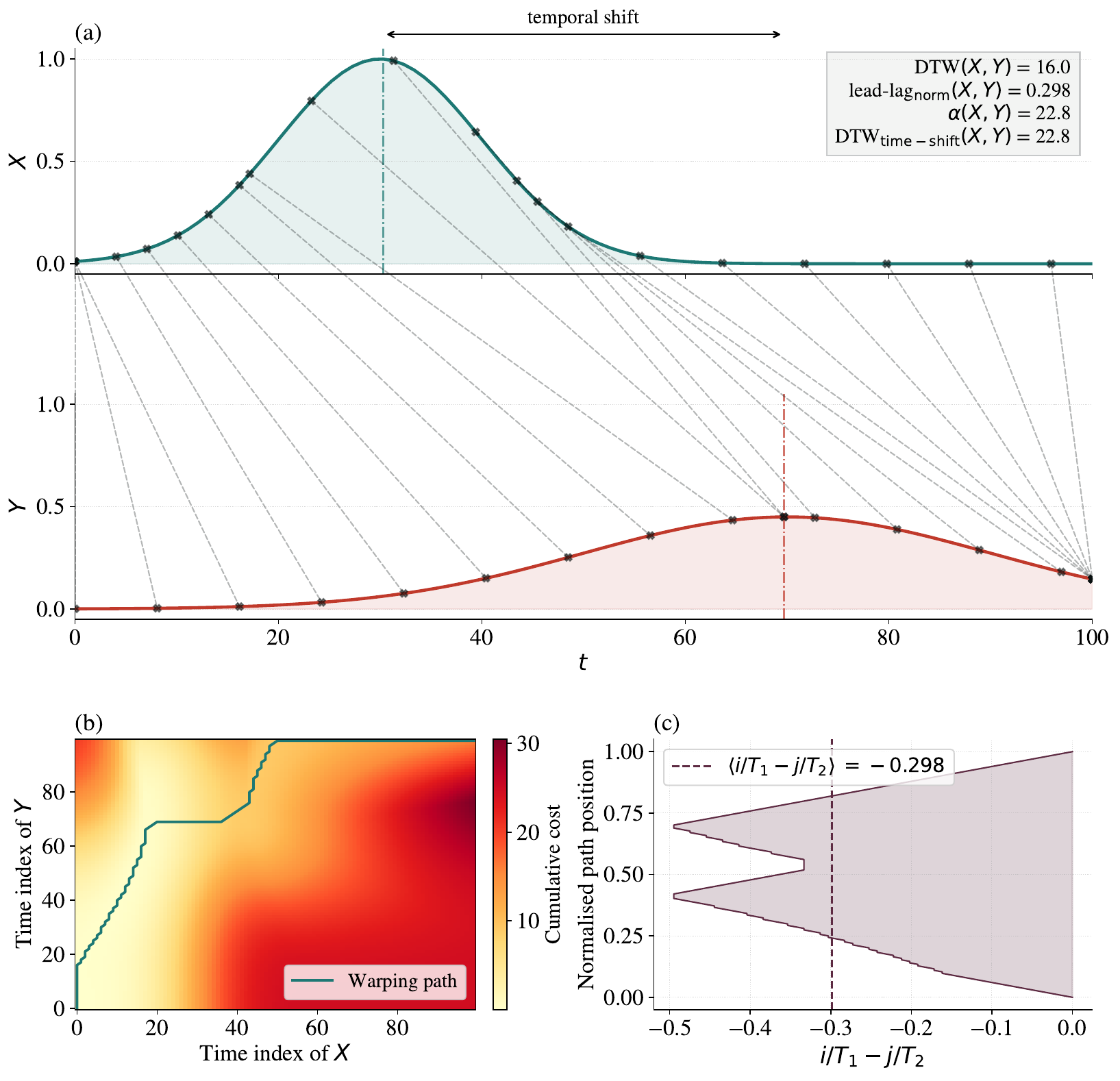}
    \caption{\textbf{Visual representation of the Time-Shift DTW.} \textbf{(a)} DTW alignment between two trajectories. The dashed gray lines illustrate the point-to-point matching. \textbf{(b)} The cumulative cost matrix with the optimal warping path $p^\ast$. \textbf{(c)} The normalized index difference $i/T_1 - j/T_2$ evaluated along the normalized path position. The vertical dashed line indicates the average of these differences, yielding the normalized lead-lag, used to penalize the temporal misalignment.}
    \label{fig:TSDTW}
    \end{figure}
Given two time series $X=\{x_t\}_{t=1,\dots,T_1}$ and $Y=\{y_t\}_{t=1,\dots,T_2}$, the optimal DTW warping path $p^\ast(X,Y)$ associates each element of $X$ with at least one element of $Y$. We define the normalized lead--lag as the average difference between the aligned time indices along the warping path:
\begin{align}
	\label{eq:norm_lead_lag}
	\text{lead-lag}_{\text{norm}}(X,Y) = \left| \frac{1}{L} \sum\limits_{(i, j) \in p^\ast(X,Y)} \left( \frac{i}{T_1} - \frac{j}{T_2} \right) \right|\,\,.
\end{align}
The \emph{Time-Shift DTW} is a linear combination of the DTW cost function and the normalized lead-lag:
\begin{align}
    \text{DTW}_{\text{time-shift}}(X,Y) = \text{DTW}(X,Y) + \alpha(X,Y) \cdot \text{lead-lag}_{\text{norm}}(X,Y)\,\,,
    \label{eq:time_shift_dtw}
\end{align}
where $\alpha(X, Y) = \frac{1}{2}\left(\sum_{n=1}^{T_1} x_n + \sum_{m=1}^{T_2} y_m\right)$ balances the two terms appearing in Eq.~\eqref{eq:time_shift_dtw}. Figure~\ref{fig:TSDTW} gives a visual interpretation of the \emph{Time-Shift DTW}. Section \ref{sec:methods:tsdtw} provides additional details on the derivation and validation of this dissimilarity measure.

\medskip

\begin{figure}[t!]
    \centering
    \includegraphics[width=\linewidth]{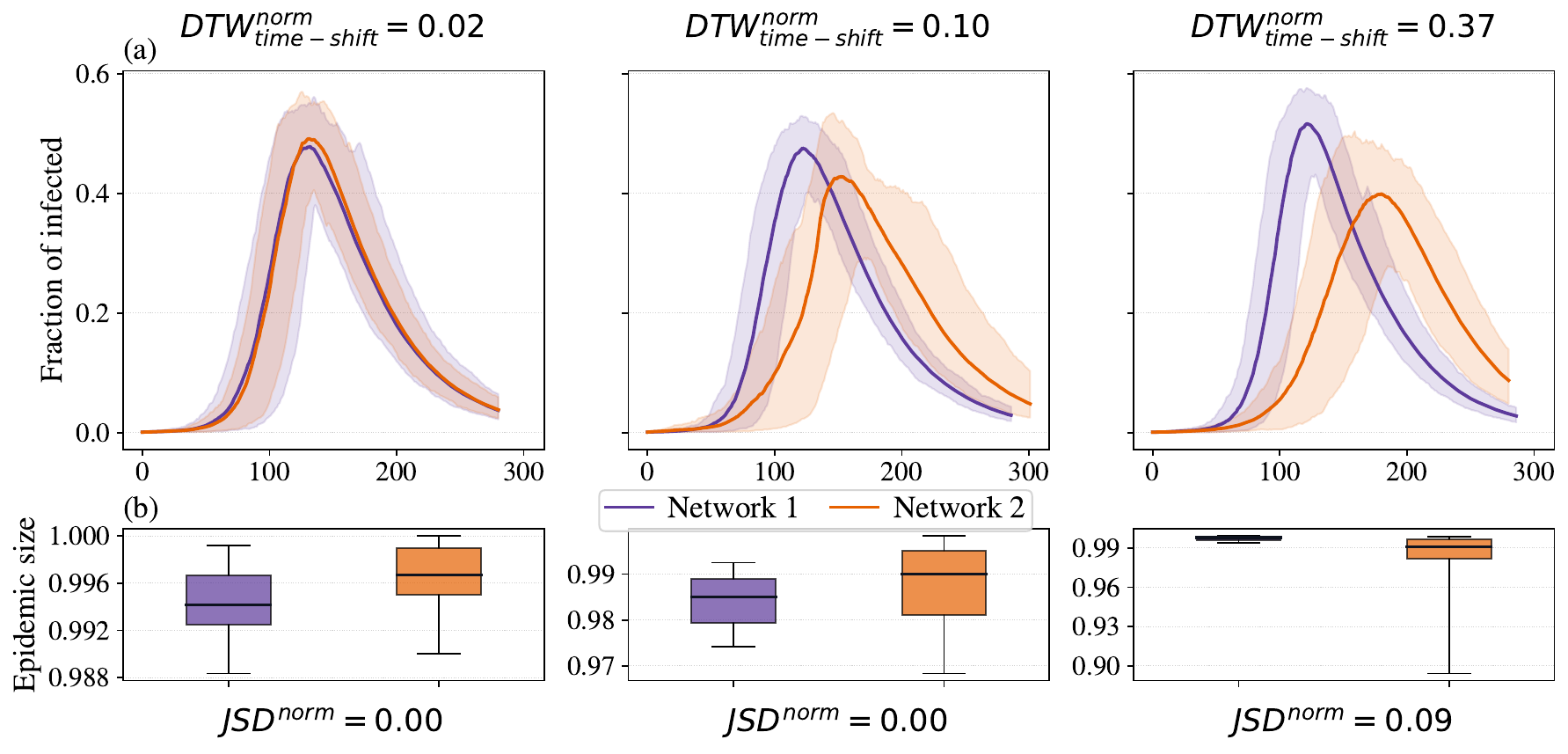}
    \caption{\textbf{Comparison of epidemic similarities across three pairs of curves of number of infected individuals.} \textbf{(a)}
    	Epidemic trajectories, represented by the fraction of infected nodes over time, for three pairs of temporal networks (Network 1 and Network 2 in each pair). The details of the implementation are given in the Methods. Solid lines indicate the mean trajectories, while the shaded areas delimit the 5th--95th percentile bands of the simulations. The curves display increasing levels of dynamic discrepancy and temporal misalignment from left to right, yielding $\text{DTW}^{\text{norm}}_{\text{time-shift}} = 0.02$, $0.10$, and $0.37$, respectively. Both dissimilarity measures (${\rm DTW}^{\rm norm}_{\rm time-shfit}$ and ${\rm JSD}^{\rm norm}$) are normalized by their respective 90th percentiles across the dataset, allowing their values to be compared on the same scale; normalized quantities are denoted by the superscript $\mathrm{norm}$. \textbf{(b)} Boxplots of the corresponding final epidemic sizes. All three pairs yield highly similar final distributions, with $\text{JSD}^{\rm norm} = 0$, $0$, and $0.09$ for the left, center, and right panels, respectively.}
    \label{fig:comparison}
\end{figure}

We compare the \emph{Time-Shift DTW} with the commonly adopted Jensen-Shannon divergence (JSD) between the distributions of epidemic sizes. To enable comparison between the two measures on the same scale, we normalize each by its respective 90th percentile across the considered dataset, denoting the resulting quantities by the superscript $\mathrm{norm}$. In Figure~\ref{fig:comparison}, we consider three pairs of curves (extracted from the ensemble of simulations detailed in Section~\ref{sec:methods.epi}, using the synthetic dataset shown in the Supplementary Material) that yield similar epidemic sizes ($\text{JSD}^{\rm norm} = 0$, $0$, and $0.09$) but have different profiles and lead to varying values of the \emph{Time-Shift DTW}. For the first pair of curves (left), the epidemic trajectories are almost identical, resulting in a near-zero score for both metrics. However, in the subsequent examples (center and right), the curves exhibit increasing temporal misalignment, with trajectories showing shifted and dilated peaks. The JSD fails to capture these dynamic discrepancies. In contrast, the \emph{Time-Shift DTW} assigns an increasing dissimilarity score, as expected ($\text{DTW}^{\rm norm}_{\text{time-shift}} = 0.02, 0.10, 0.37$).

\subsection{Epidemic-Aware Temporal Graph Dissimilarity}

The \emph{Time-Shift DTW} can effectively compare epidemic curves. However, using it to assess the dissimilarity between temporal graphs in terms of epidemic response requires simulating multiple realizations of the spreading process on each network, resulting in a high computational cost. We thus introduce a notion of temporal graph dissimilarity, depending on the behavior of a spreading process unfolding on the graphs under comparison. This quantity is calibrated using stochastic simulations to reflect dissimilarity in terms of the \emph{Time-Shift DTW}, but, once calibrated, it is a deterministic and differentiable quantity that can be used as a loss function. For the stochastic simulations, we consider an SIR model with infection and recovery parameters chosen to exceed the theoretical epidemic threshold for temporal graphs \cite{valdano2018epidemicthreshold}, ensuring the spread of the pathogen (additional details are provided in Section \ref{sec:methods.epi}). We choose SIR for having  fewer parameters than more realistic epidemic models (\emph{e.g.}, SEIR, SIRD), while producing remarkably similar dynamical curves as shown in Ref.~\cite{contreras2024infection}.

\subsubsection{Definition of the dissimilarity measure}

Our dissimilarity measure consists of two terms. The first is
the embedding-based distance for temporal graphs proposed in Ref.~\cite{Dall_Amico_2024}. This metric first obtains a node embedding for each graph that captures node similarity based on time-respecting random walks \cite{dallamico2025learning}, and then computes the graph distance in the embedding space. Since time-respecting paths drive diffusion processes in temporal networks, we expect this distance to capture relevant aspects of epidemic dynamics.

The second term is a combination of graph feature corrections whose weights are trained to improve the epidemic sensitivity of the baseline distance. Following Ref.~\cite{longa2024generating}, we consider a set of $15$ graph features that are relevant to spreading on temporal networks. We group them into topological (\ie properties of the time-aggregated network) and temporal (pertaining to the dynamical evolution of interactions over time) categories, as summarized in Table~\ref{tab:graph_features}. 
  \begin{table}[t!]
	\centering
	\footnotesize                                                                                     
	\setlength{\extrarowheight}{2pt}
	\begin{tabularx}{\columnwidth}{@{} l >{\itshape}l X @{}}                  
		\toprule                                                                               
		\textbf{Category} & \normalfont{\textbf{Name}} & \textbf{Definition} \\
		\midrule
		\multirow{6}{*}{\emph{Topological}}
		& avg\_strength        & Average node strength \\
		& std\_strength        & Standard deviation of strength across nodes \\                
		& weighted\_clustering         & Average fraction of closed triangles per node \\
		& assortativity & Degree assortativity coefficient \\
		& community\_entropy & Quantifies the strength of a community structure \\
		& n\_nodes           & Number of nodes \\
		\midrule
		\multirow{9}{*}{\emph{Temporal}}
		& avg\_edges\_per\_t   & Average number of active edges per time step \\
		& std\_edges\_per\_t   & Standard deviation of active edges per time step \\
		& burstiness    & $(\sigma - \mu)/(\sigma + \mu)$, where $\mu,\sigma$ are the mean and std.\ of
		inter-event times \\
		& avg\_lifetime\_edges & Average duration of a temporal edge \\
		& std\_lifetime\_edges & Standard deviation of temporal edge duration \\
		& avg\_edge\_stability & Average Jaccard similarity between consecutive snapshots \\
		& std\_edge\_stability & Standard deviation of edge stability \\
		& avg\_activity        & Fraction of time steps in which a node has at least one neighbour \\
		& std\_activity        & Standard deviation of node activity \\
		\bottomrule
	\end{tabularx}
	\caption{\textbf{Topological and temporal graph features used in the analysis.} Detailed definitions are provided in Section~\ref{sec:methods.features}.}                        
	\label{tab:graph_features}
\end{table}
\begin{table}[b!]
    \centering
    \small                                                                              
    \setlength{\extrarowheight}{2pt}
    \begin{tabularx}{\columnwidth}{@{} l X c c @{}}                  
        \toprule                                                                                
        \textbf{Context} & \textbf{Description} & \textbf{Nodes} & \textbf{Duration} \\
        \midrule
        Primary School \cite{school_dataset_3} & Students and teachers from a primary school & 242 & 2 days \\
        High School 1 \cite{school_dataset_2} & Students and teachers from a high school & 126 & 4 days \\
        High School 2 \cite{school_dataset_2} & Students from a high school & 180 & 7 days \\
        High School 3 \cite{school_dataset} & Students from a high school & 327 & 5 days \\
        Conference \cite{workplace_dataset_2} & Participants in a scientific conference & 405 & 2 days \\
        Office \cite{workplace_dataset_2} & Workers in an office & 232 & 10 days \\
        Hospital \cite{hospital_dataset} & Patients and health-care workers in a hospital & 75 & 5 days \\
        Malawi \cite{malawi} & Individuals in a village in rural Malawi & 86 & 14 days \\
        \bottomrule
    \end{tabularx}
    \caption{\textbf{Summary description of the SocioPatterns social networks.} It includes a description of the environment, the number of nodes and the duration of the data collection.}
    \label{tab:empirical_datasets}
\end{table}
Topological features characterize the structure of the time-aggregated network and include quantities such as the strength distribution -- where a node's strength is defined as the sum of the weights of its edges -- the weighted clustering coefficient, and degree assortativity \cite{newman_networks, barabasi2013network}. Temporal features, in turn, capture the evolution of contacts over time. In particular, we include measures known to be relevant to epidemic spreading, such as the burstiness of contact events \cite{holme_saramaki, karsai_bursty}, the heterogeneity of node activity \cite{perra2012activity}, and the temporal distribution of active edges \cite{starnini2012random}. We define the \acronym as follows:
\begin{equation}
    \text{\acronym}(\mathcal{G}_1, \mathcal{G}_2) = \log \left[d(\mathcal{G}_1, \mathcal{G}_2) +1 \right] + {\sum_i\gamma_i \left[f_i(\mathcal{G}_1)- f_i(\mathcal{G}_2)\right]^2}\,\,,
    \label{proposal}
\end{equation}
where $d(\mathcal{G}_1, \mathcal{G}_2)$ is the unmatched temporal graph distance of Ref.~\cite{Dall_Amico_2024} between graphs $\mathcal{G}_1$ and $\mathcal{G}_2$, $f_i(\mathcal{G})$ denotes the $i$-th graph feature (\emph{e.g.}, the clustering coefficient), while $\gamma_i$ is a learnable parameter controlling the contribution of the corresponding corrective term. The parameters $\gamma_i$ are calibrated on a training set of temporal graphs to increase the correlation with the ground truth \textit{Time-Shift DTW}. Each feature $f_i$ is standardized to zero mean and unit variance using statistics computed exclusively on the training set. The distance $d$ has the advantage of not requiring any training and can be evaluated for any pair of temporal graphs. However, as we will show in the following, it may over- or under-weight some features in terms of epidemic spreading. The learnable weights $\gamma_i$ allow us to correct the baseline graph distance to better match the \emph{Time-Shift DTW}.

\subsubsection{Calibration on empirical data}

\begin{figure}[t!]
	\centering
	\includegraphics[width=\linewidth]{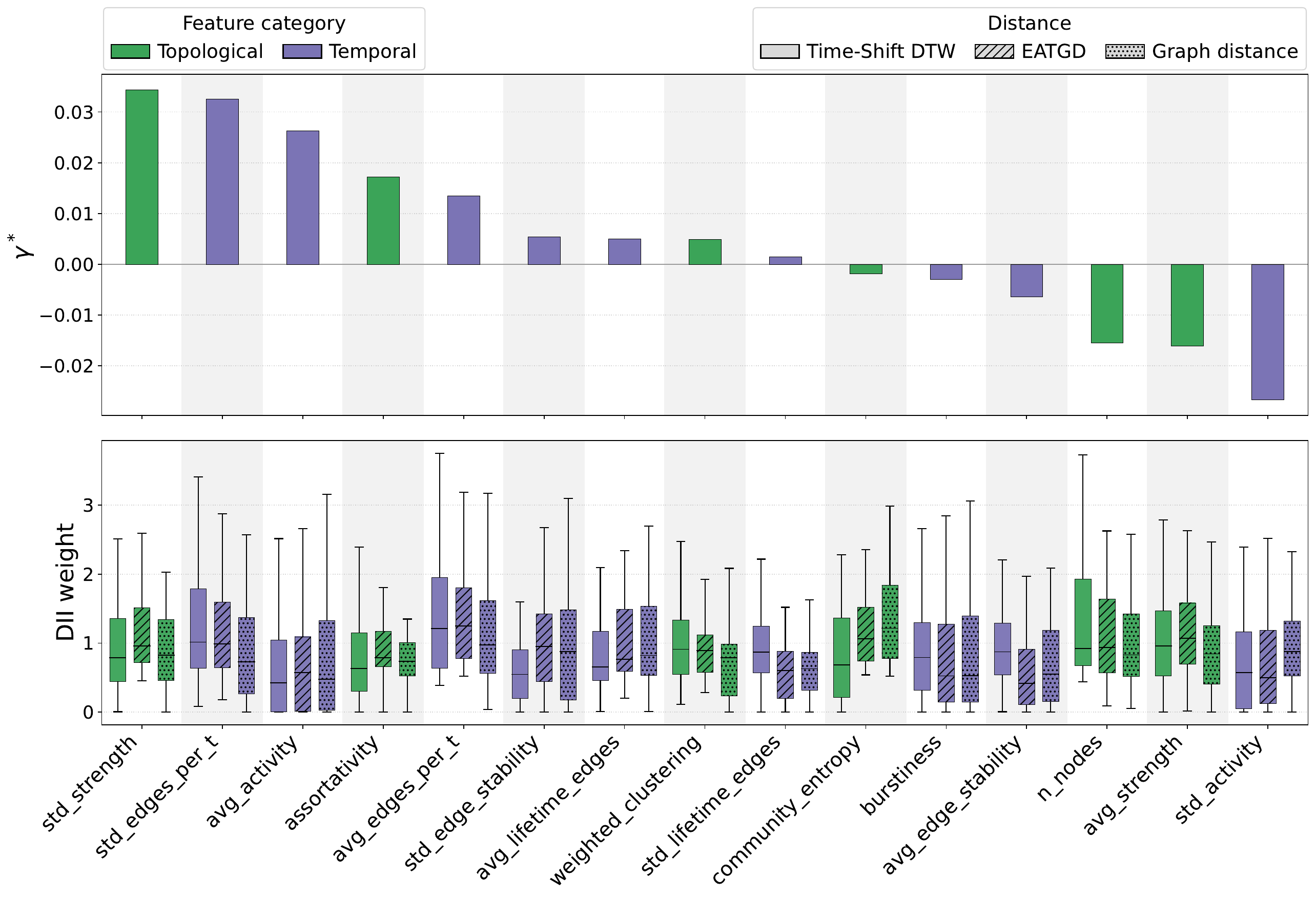}
	\caption{\textbf{Feature importance for the calibrated EATGD.} (Top) Learned weights $\gamma_i^*$ assigned to each of the $15$ graph features when calibrating EATGD on the SocioPatterns training set, colored by feature category (topological vs. temporal). (Bottom) Distribution of the importance weights attributed by the Differentiable Information Imbalance (DII) to each feature, for three target dissimilarities: Time-Shift DTW, the baseline temporal graph distance, and the optimized EATGD. Boxplots are obtained from $50$ independent DII runs with random initializations.}
	\label{fig:optimized_gammas_sp}
\end{figure}
We calibrate the parameters $\boldsymbol{\gamma} = (\gamma_1, \dots, \gamma_F)$ on empirical proximity data collected by the SocioPatterns collaboration \cite{cattuto2010dynamics}. These temporal networks are measured using wearable proximity sensors and encode high-resolution face-to-face proximity interactions within approximately $1.5$ meters. Table~\ref{tab:empirical_datasets} provides a brief summary of these datasets. Each data collection spans multiple days, and we treat each day as a separate temporal graph. The parameters $\bm{\gamma}$ are learned by running multiple SIR simulations on the analyzed graphs and then solving the following optimization problem:
\begin{equation}
	\boldsymbol{\gamma}^* = \arg\min_{\boldsymbol{\gamma}} \left[-\rho \left( \mathbf{D}'(\boldsymbol{\gamma}), \mathbf{D}_{\text{epi}} \right) + \lambda||\boldsymbol{\gamma}||^2\right]\,\,,
	\label{correlation_formula}
\end{equation} 
where $\rho$ denotes the Mantel correlation coefficient, $\mathbf{D}'(\boldsymbol{\gamma})$ denotes the pairwise graph dissimilarity matrix obtained using 
\acronym and $\mathbf{D}_{\text{epi}}$ is computed using the \emph{Time-Shift DTW}.

We evaluate \acronym on a test set, composed of one randomly selected day from each dataset, after calibrating the parameters $\bm{\gamma}$ on a training set, containing the remaining graphs. The initial correlation $\rho \approx 0.42$ between the graph distance and the \emph{Time-Shift DTW} increases after training to $\rho \approx 0.83$ on the training set and $\rho \approx 0.79$ on the test set, showing the ability of \acronym to faithfully capture epidemic dissimilarity. The top plot of Figure~\ref{fig:optimized_gammas_sp} shows the learned parameters $\gamma_i$, displaying the relative importance of each feature in achieving the alignment between the two dissimilarity measures.

\medskip

We further evaluate the relationship between \acronym\ and the \emph{Time-Shift DTW} using the Differentiable Information Imbalance (DII) \cite{weight_dadapy} on the training set to identify the features driving the alignment between the two measures. DII automatically estimates feature importance by learning the optimal weights that best align a feature-based distance with a target dissimilarity measure. 
Specifically, we use DII to quantify the contribution of each graph feature to the \emph{Time-Shift DTW} and then repeat the same analysis using the temporal graph distance of Ref.~\cite{Dall_Amico_2024} and the optimized \acronym as target dissimilarities. For each dissimilarity measure, similarly to Ref.~\cite{bioglio2016recalibrating}, we perform $50$ independent optimization runs with different random initializations, obtaining a distribution of optimal feature weights. We compare these distributions using Cohen's $d$ effect size \cite{cohen2013statistical}.
As illustrated in the second row of Figure~\ref{fig:optimized_gammas_sp}, the relative importance of the different network features remains consistent across all three measures. Remarkably, we observe that the total Cohen's $d$ (obtained by summing Cohen's $d$ over all features) decreases from $d = 3.3$ to $d = 3.13$ after training, indicating greater agreement with the \emph{Time-Shift DTW} also according to this analysis.

\medskip

We note that the coefficients learned in this scenario reflect characteristic properties of real-world close-range human interactions and capture aspects of the physical and temporal constraints governing human spatial proximity. As a result, they provide a natural pre-trained baseline for applications involving proximity networks with characteristics similar to those of the SocioPatterns datasets. 
The framework itself, however, is not tied to a specific application domain. By repeating the optimization procedure on representative data, the measure can be adapted to temporal networks generated by different interaction mechanisms, such as online social platforms or communication networks. As such, in the Supplementary Material, we evaluate the framework on synthetic temporal graphs, demonstrating its applicability beyond proximity networks.
\begin{figure}[t!]
	\centering
	\includegraphics[width=\textwidth]{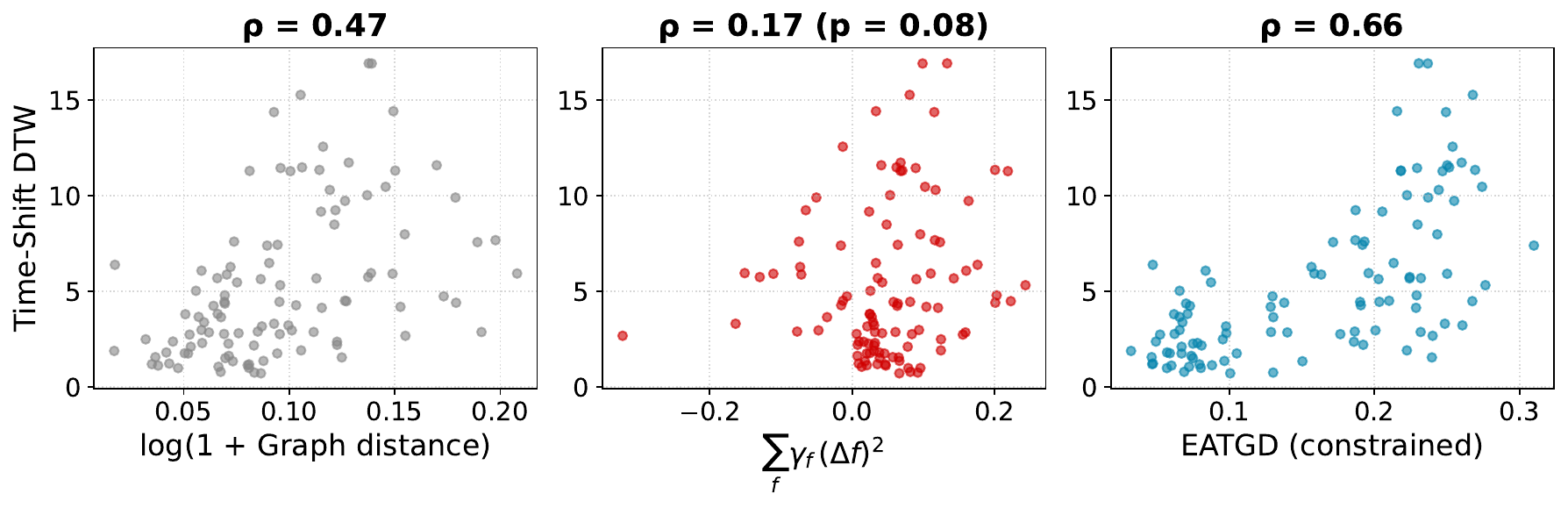}
	\caption{\textbf{Performance of EATGD in out-of-distribution scenarios.} The scatter plots compare the \emph{Time-Shift DTW} with: (left panel) the baseline temporal graph distance, $\log[d(\mathcal{G}_1,\mathcal{G}_2)+1]$; (center panel) the unconstrained feature-based component of EATGD, $\sum_f \gamma_f(\Delta f)^2$, excluding the graph-distance term; and (right panel) the constrained EATGD, which includes only the features that are well represented in the training set, as described in the main text. The titles report the Pearson correlation coefficient $\rho$. The $p$-values are reported only for non-significant correlations.}
	\label{fig:gating_example}
\end{figure}

\subsubsection{Robustness of the measure}

By design, \acronym requires a training set for calibration, and the learned weights $\bm{\gamma}$ generally depend on the training data, potentially limiting its applicability to previously unseen patterns. In this section, we show that the graph-distance component of \acronym improves its robustness and allows it to generalize to unseen contexts. Indeed, this component requires no training and, as shown above, exhibits a significant correlation with the \emph{Time-Shift DTW}.

\medskip

To verify this hypothesis, we designed a different experimental setting so that all graphs in the test set were collected in settings (\emph{e.g.} schools, offices, villages) not present in the training set. This design choice allows us to consider cases in which the training set is not representative of the test set and the learned weights $\bm{\gamma}$ may be detrimental to the overall performance of the dissimilarity.

We then introduce a statistical gating mechanism to train \acronym: each feature is included only if its value for the test graph is well represented in the training set. In practice, we require that the feature value deviates by at most two standard deviations from the training-set mean.  In the worst-case scenario, \acronym reduces to $\log \left[d(\mathcal{G}_1, \mathcal{G}_2) +1 \right]$, independently of the learned coefficients. Conversely, when the test graph is very representative of the training set, all feature terms are included, thereby retaining the full feature correction. This ensures that \acronym remains adaptable to in-distribution graphs while maintaining reliability in unfamiliar domains. Figure~\ref{fig:gating_example} shows a representative cross-environment evaluation using the \emph{Highschool 2012} and \emph{Malawi} datasets as test sets. The scatter plots compare the \emph{Time-Shift DTW} with only the distance-based component of \acronym (first panel), the feature-based component of \acronym calibrated using all features and without the distance-based term (second panel), and the constrained \acronym detailed above. When relying only on graph features, the inclusion of out-of-distribution feature values reduces the alignment with the \emph{Time-Shift DTW}. The constrained \acronym selectively excludes correction terms associated with out-of-distribution feature values. This gating improves the correlation and outperforms the graph-distance baseline. Additional scatter plots for other dataset pairs are provided in the Supplementary Material.

\section{Discussion}

Evaluating the similarity between temporal graphs is an essential task in network science. Common approaches \cite{wills2020metrics,tantardini2019comparing, sanfeliu1983distance, koutra2013deltacon, berlingerio2012netsimile, bagrow2019information, tsitsulin2018netlsd} compare graphs based on their topological properties, \ie by analyzing how connections are distributed across nodes. These approaches, however, can be highly sensitive to differences in features that may not influence the outcome of dynamical processes unfolding on the network, such as epidemic spreading. We thus devise a dissimilarity measure capturing how much two graphs generate comparable spreading dynamics. To achieve this, we exploit a targeted set of topological and temporal features specifically tuned to capture epidemic outcomes. This perspective requires less information: structurally different graphs may induce nearly indistinguishable epidemic outcomes, meaning that epidemic trajectories capture the relevant functional signal without requiring an exact reproduction of the underlying network structure.

\medskip

We first introduce a notion of graph dissimilarity based on epidemic simulations. This is done by running multiple realizations of the SIR model on pairs of temporal graphs and comparing the resulting epidemic curves using the \emph{Time-Shift DTW}, a novel dissimilarity measure that is sensitive both to the shape of the epidemic trajectory and to global temporal delays. However, this strategy is computationally expensive because it relies on repeated stochastic simulations. We therefore develop a differentiable dissimilarity measure, which we call \name (\acronym) that combines an embedding-based temporal graph distance with topological and temporal graph descriptors, whose contributions are optimized to maximize the correlation with the \emph{Time-Shift DTW}. In this way, \acronym provides an efficient, deterministic, and differentiable proxy for epidemic-based graph dissimilarity. We show the effectiveness of this approach on empirical datasets, using close-range proximity networks from the SocioPatterns collaboration. Despite the heterogeneity of empirical human interaction patterns, we empirically verified the ability of \acronym in capturing the underlying spreading dynamics. Moreover, extensive experiments on a large-scale dataset of synthetic temporal graphs demonstrate the robustness and scalability of the proposed pipeline.

\medskip

Our work entails some limitations worth discussing. First, although the validation explores a broad range of topological and temporal properties, the considered graph space remains inherently limited. Consequently, the trained parameters defining \acronym depend on the training data considered. Second, \acronym is calibrated against the \emph{Time-Shift DTW}, which itself depends on the epidemic model adopted for the simulations. In this work, we considered an SIR process whose transmission and recovery probabilities were selected such that the process operates above the theoretical epidemic threshold \cite{valdano2018epidemicthreshold}. Although the SIR model produces similar dynamical curves to the SEIR and SIRD models \cite{contreras2024infection}, alternative and more complex propagation models, as well as different epidemic regimes, could affect the resulting dissimilarity landscape. Third, once calibrated, \acronym is deterministic, differentiable, and computationally efficient, making it suitable for evaluating generative models or serving as a loss function for their training. However, the calibration step itself remains computationally intensive, as it requires estimating the weights from multiple realizations of the SIR model. Importantly, this calibration only needs to be performed once for a given setting.

\medskip 

Despite these limitations, the proposed framework remains highly flexible. The parameters of \acronym can be retrained, as demonstrated using the empirical and synthetic datasets, for different spreading rates, and the measure can naturally incorporate additional graph descriptors whenever needed.

Our findings also open several promising research directions. Although we focused on epidemic spreading, the same framework could naturally be extended to other dynamical processes shaped by network structure, such as synchronization \cite{barrat2008dynamical}. 

The epidemic-based perspective developed here is well suited for the evaluation and training of generative models of temporal networks \cite{primer_on_tgnn, longa2023graph, survey_tgnn, taggen}, which aim to produce realistic surrogate temporal networks by learning distributions of temporal edges directly from data. Current generative approaches \cite{primer_on_tgnn, longa2023graph, survey_tgnn, taggen} are typically optimized to minimize topological reconstruction errors, an objective that requires large training datasets, which are unavailable in most empirical scenarios, to capture the full structural complexity of the graphs. 

Through extensive experiments on empirical human proximity datasets from SocioPatterns and on synthetic temporal graphs, we show that EATGD accurately captures epidemic dissimilarity. Because EATGD is deterministic and differentiable once calibrated, it avoids stochastic epidemic simulations at inference time, opening the possibility of using it as a loss function to train generative models that produce temporal graphs with realistic epidemic dynamics without necessarily replicating the fine-grained structural details of the training data. This could enable a more data-efficient training paradigm, allowing the generation of epidemiologically valid surrogate networks even when empirical data is scarce.

Finally, the present framework measures global dissimilarity between unmatched temporal graphs, where no correspondence exists between the nodes of the compared networks. A natural extension would be to incorporate node-level attributes — such as age, occupation, or spatial location — directly into both the epidemic dynamics and the resulting dissimilarity measure. Such attributed temporal graphs arise naturally in epidemiological and public-health applications, where demographic and geographical heterogeneity strongly influence transmission patterns. Extending \acronym to this richer setting represents a straightforward and promising generalization of the current methodology.

\section{Methods}

\subsection{Time-Shift DTW}
\label{sec:methods:tsdtw}

\subsubsection{Derivation}

Dynamic Time Warping evaluates morphological similarities but remains invariant to global temporal translations. However, when comparing epidemic curves, shifts are important because they represent delays in the infection time. We hence introduce a penalty term to account for shifts explicitly. We take the \emph{lead-lag} \cite{lead_lag} as a starting point for our correction term. Given the optimal warping path $p^*$, it is quantified as the mean difference of the indices in $p^*$,
\begin{align*}
\text{lead-lag}(X,Y) = \left|\frac{1}{L} \sum_{(i,j)\in p^*(X,Y)}(i-j) \right|\,\,.
\end{align*}
However, this quantity strongly depends on the length of the time series under comparison.  To allow a fair comparison between time series of different duration, we divide the path indices $i$ and $j$ by their respective sequence lengths, $T_1$ and $T_2$. Averaging the local relative temporal shifts $(i/T_1 - j/T_2)$ over the entire warping path of length $L$ yields the $\text{lead-lag}_{\rm norm}(X,Y)$ defined in Eq.~\ref{eq:norm_lead_lag}. This normalization maps the temporal penalty onto the interval $[0, 1]$. We then introduce the scaling factor $\alpha(X,Y)$ to properly scale the temporal penalty and define it as the average cumulative sum of the two epidemic trajectories:
\begin{align*}
	 \alpha(X,Y) = \frac{1}{2} \left( \sum_{n=1}^{T_1} x_n + \sum_{m=1}^{T_2} y_m \right)\,\,.
\end{align*}
This formulation ensures that the temporal penalty scales proportionally with the epidemic size. By multiplying the normalized shift by $\alpha$, a temporal delay in a severe epidemic incurs a proportionally higher absolute penalty than an equivalent delay in a minor outbreak, providing an epidemiologically motivated balance between shape distortion and temporal misalignment.

\subsubsection{Evaluation}

\begin{figure}[!t]
	\centering
	\includegraphics[width = 0.7\columnwidth]{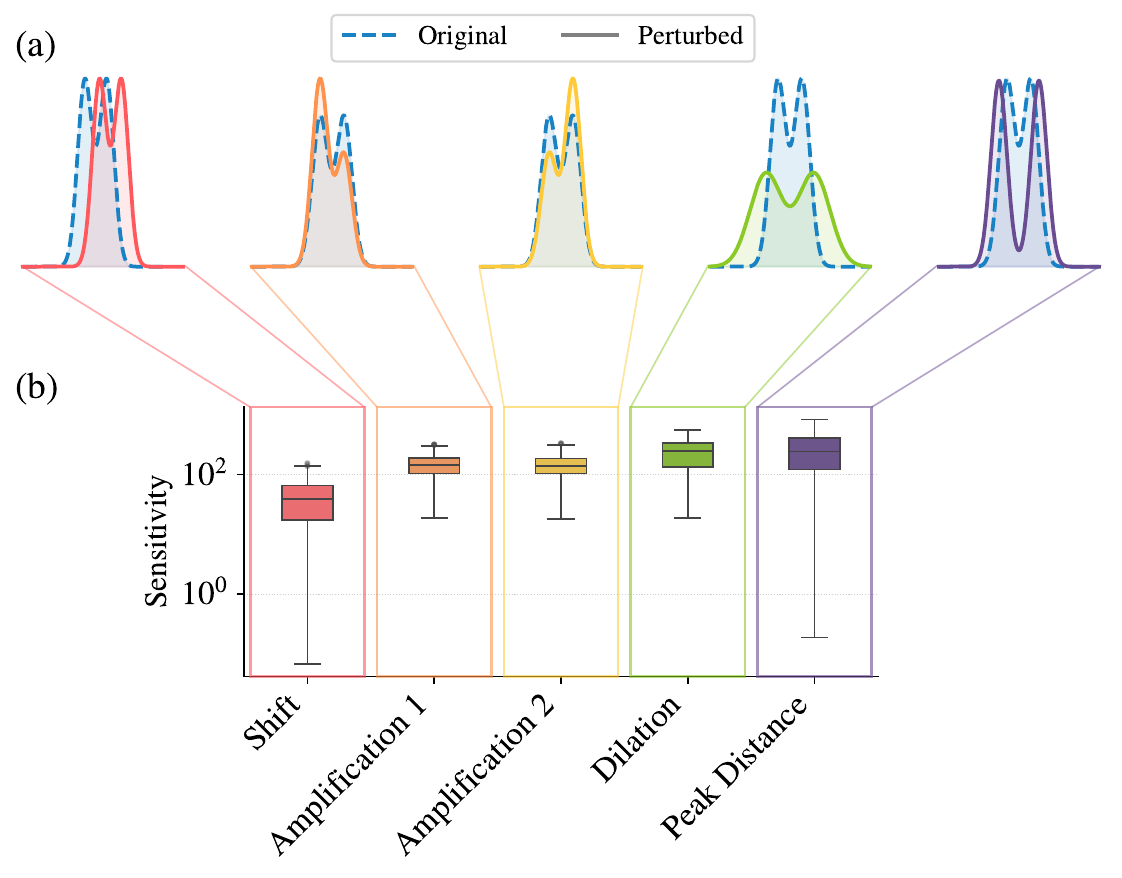}
	\caption{\textbf{Sensitivity analysis of the Time-Shift DTW.} \textbf{(a)} Schematic representation of the morphological transformations applied to the baseline bi-Gaussian curve (dashed blue lines) to generate the perturbed trajectories (solid outlines). \textbf{(b)} Distributions of the empirically computed sensitivities for each parameter. The strictly positive boxplots, evaluated over a large set of randomly sampled configurations, demonstrate the measure's consistent responsiveness to diverse temporal and morphological variations.}
	\label{fig:gradients_plot}
\end{figure}

We evaluate the definition of \emph{Time-Shift DTW} by verifying that is sensitive to differences in peak position, height, width, or number of peaks. Specifically, we consider a synthetic epidemic curve $h(t) = h_1(t) + h_2(t)$, obtained as the sum of two Gaussian functions with means $\mu_1, \mu_2$ and variances $\sigma_1^2, \sigma_2^2$, respectively. We then consider four transformations that modify $h(t)$ into $g(t)$ by altering the peak heights, positions, and widths. These transformations are listed below and illustrated in Figure~\ref{fig:gradients_plot}a.
\begin{itemize}
	\item Shift: A global translation $s$ of the entire curve along the time axis, representing an overall delay in the outbreak onset: $g(t) = h(t-s)$
	\item Peak amplification: Two independent scaling factors, $a_1, a_2$, control the amplitudes of the two peaks and simulate different severities of the primary and secondary infection waves: $g(t) = a_1 h_1(t) + a_2 h_2(t)$.
	\item Dilation: The standard deviations are multiplied by a factor $\theta$, broadening the peaks and representing slower, more prolonged spreading dynamics: $\sigma'_1~=~\theta \sigma_1, ~\sigma'_2 = \theta \sigma_2$.
	\item Peak Distance: The temporal separation between the centers of the two peaks $\Delta \mu = |\mu_2-\mu_1|$, capturing the temporal separation between successive epidemic waves: $g(t) = h_1(t-\Delta \mu/2 ) + h_2(t + \Delta \mu/2)$.
\end{itemize}
	We first sample a large number of curves $g(t)$ by randomly  combining these transformations. We then rescale the curves of infected individuals by a multiplicative factor to preserve the area under the curve, which is proportional to the epidemic size. Finally, we quantify the measure's sensitivity to each transformation by applying a small perturbation and computing the derivative numerically via finite differences. We quantify the \emph{sensitivity} as the product of the gradient and the parameter value, enabling a dimensionless comparison across variables with heterogeneous magnitudes.

Figure~\ref{fig:gradients_plot}b shows the distributions of the sensitivities, which are strictly positive for all evaluated transformations, confirming the goodness of the definition of our dissimilarity measure.

\subsection{Empirical temporal graphs}
\label{sec:methods.tgraphs}

The proposed framework is evaluated on eight real-world human contact networks from the SocioPatterns collaboration \cite{cattuto2010dynamics}. These datasets collect high-resolution proximity contact data across various environments: primary and secondary schools \cite{school_dataset, school_dataset_2, school_dataset_3}, a workplace \cite{workplace_dataset_2}, a healthcare facility \cite{hospital_dataset}, an academic conference \cite{workplace_dataset_2}, and a rural community in Malawi \cite{malawi}. 

Data collection relies on wearable proximity sensors that exchange radio-frequency packets within a spatial range of 1 to 1.5 meters, enabling the detection of relevant social interactions under this threshold. In particular, the system records an active contact event whenever two devices exchange at least one radio packet within a 20-second time interval, yielding a high-resolution temporal mapping of the interaction dynamics.
Table \ref{tab:empirical_datasets} outlines the main characteristics of each dataset.

\subsection{Epidemics simulations}
\label{sec:methods.epi}

For each graph, we simulate epidemic spreading using the SIR model on the temporal graph \cite{1927sir}. We fix the recovery probability at $\mu = 0.02$ and consider five values of the transmission probability, $\beta \in \{0.15, 0.20, 0.25, 0.30, 0.35\}$, in order to cover spreading scenarios with different speeds. These values are chosen to ensure that the epidemic threshold \cite{valdano2018epidemicthreshold} is consistently exceeded, falling within a realistic regime for modeling respiratory diseases as established in related literature \cite{dyane, cencetti2025generating, sapienza2018estimating}.

For each combination of graph and $\beta$, we run 50 independent realizations, each seeded from a single randomly chosen node. To ensure statistical relevance, we retain only epidemics that ``take off'', defined as outbreaks in which at least 10\% of the population is infected by the end of the process.

\subsection{Graph Features Definitions}
\label{sec:methods.features}

We characterize the structural and dynamic properties of the networks using a specific set of topological and temporal features. These quantities capture multiple characteristics of the temporal graphs, ranging from the macroscopic organization of the network (community structure) to the microscopic temporal dynamics of individual contacts (burstiness, stability, and node activity).

\paragraph{Community structure.}

The community structure of a graph refers to the tendency of nodes to organize into densely connected groups with sparser connections between them. We quantify the presence and strength of such modular organization in the aggregated network via the \emph{community entropy} using the embedding computed through the temporal graph distance \cite{Dall_Amico_2024} algorithm. Let $X \in \mathbb{R}^{n \times d}$ be the node embedding matrix, $n$ number of nodes in the graph and $d$ the embedding dimension. We construct the normalized covariance matrix $\frac{X^TX}{n}$.
Since this matrix is positive semi-definite, its eigenvalues are real and non-negative. In the presence of $k$ communities, the embedding $X$ is well approximated by a mixture of $k$ Gaussians~\cite{dallamico2025learning}. From Weyl's inequality~\cite{bhatia2013matrix}[Corollary III.2.2], the top $k-1$ eigenvalues of the covariance matrix converge to the eigenvalues of the expectation $\mathbb{E}\left[\frac{X^TX}{n}\right]$, which are determined by the mixture means. These $k-1$ eigenvalues are \emph{isolated}, \ie they are much larger than the rest. We thus compute an entropy measure of these eigenvalues to assess the presence of a community structure in the graph, by defining the probability distribution $\mathbf{p} = \{p_1, \dots, p_d\}$, where each component is 
\begin{equation*}
    p_i = \frac{e_i}{\sum_{j =1}^d e_j}.
\end{equation*} 
with $e_i$ representing the $i$-th eigenvalue. The community entropy is the Shannon entropy of this distribution:
\begin{equation}
    H = -\sum_{i=1}^d p_i \ln p_i.
\end{equation}
This measure is bounded in $[0, \ln{d}]$. The presence of communities makes the probability mass $p_i$ more localized around isolated eigenvalues, making the entropy decrease.

\medskip

\paragraph{Temporal features.} The \emph{burstiness} \cite{holme_saramaki, karsai_bursty} describe the intermittent nature of the contacts by evaluating the inter-event times $\tau$ between consecutive edge activations. It is defined as: 
\begin{equation*}
    B = \frac{\sigma_\tau - \mu_\tau}{\sigma_\tau + \mu_\tau},
\end{equation*}
where $\mu_\tau$ and $\sigma_\tau$ are the mean and standard deviation of $\tau$, respectively. A value of $B > 0$ indicates a bursty, heavy-tailed contact dynamic, while $B \approx 0$ indicates a memoryless Poissonian process.

\medskip 

The \emph{edge lifetime} \cite{cattuto2010dynamics} is the continuous duration $\Delta t$ for which a temporal edge remains uninterruptedly active, while \emph{edge stability} \cite{chiappori2021quantitative} quantifies the topological memory of the network, by computing the structural similarity between consecutive temporal snapshots using the Jaccard index:
\begin{equation*}
    J_t = \frac{|E_t \cap E_{t+1}|}{|E_t \cup E_{t+1}|}.
\end{equation*}

\emph{Node activity} evaluates the empirical activity $a_i$ of a node $i$ and it is defined as the fraction of time steps in which the node is involved in at least one interaction \cite{perra2012activity}:
\begin{equation*}
    a_i = \frac{1}{T} \sum_{t=1}^T \mathbb{I}(k_{i,t} > 0),
\end{equation*}
where $k_{i,t}$ is the degree of node $i$ at time $t$, and $\mathbb{I}(\cdot)$ is the indicator function.

\subsection{Optimization hyperparameters}
The feature importance analysis using the Differentiable Information Imbalance (DII) framework was conducted by executing 50 independent runs with random initializations to ensure the statistical robustness of the derived weights. Each optimization process was run for 100 epochs using the Adam optimizer with a learning rate of $10^{-2}$.

To maximize the correlation between the \acronym and the \emph{Time-Shift DTW} (Eq.~\ref{correlation_formula}), we employed a vanilla Gradient Descent algorithm. The learning process for the $\gamma$ parameters was configured for a total of 500 epochs with a learning rate of $0.01$ and a coefficient of $\lambda = 0.01$ for the regularization term. 

\section*{Data and Code availability}

Commented codes and data to reproduce the results presented in this paper can be found at \href{https://github.com/s-freni/eatgd}{github.com/s-freni/eatgd}.

\section*{Acknowledgments}
This publication is part of the project PNRR-NGEU which has received funding from the MUR – DM 630/2024. This study was carried out within the FAIR - Future Artificial Intelligence Research and received funding from the European Union Next-GenerationEU (PIANO NAZIONALE DI RIPRESA E RESILIENZA (PNRR) – MISSIONE 4 COMPONENTE 2, INVESTIMENTO 1.3 – D.D. 1555 11/10/2022, PE00000013). This manuscript reflects only the authors’ views and opinions, neither the European Union nor the European Commission can be considered responsible for them. LD and CC acknowledge the support from the Lagrange project of Fondazione CRT.

{\footnotesize

}


\appendix

\renewcommand{\thesection}{S}
\setcounter{subsection}{0}
\setcounter{figure}{0}
\setcounter{table}{0}
\renewcommand{\thefigure}{S.\arabic{figure}}
\renewcommand{\thetable}{S.\arabic{table}}

\section*{Supplementary material}

\subsection{Validation of \acronym on the synthetic dataset}

\subsubsection{Synthetic temporal graphs generation}

We evaluate the robustness of our approach by testing it on a large dataset of synthetic temporal graphs. We generate a family of temporal graphs spanning a wide range of topological and temporal properties. We consider three generative models for static graphs as the backbone of the final temporal networks, and assign an activation time-series to each edge to obtain a temporal graph. The models are the following:
\begin{itemize}
    \item \textbf{Stochastic Block Model (SBM)} \cite{karrer2011stochastic}. This model generates graphs with a community structure. We consider $k=2$ classes and denote by $c_{\rm in}$ and $c_{\rm out}$ the within- and across-class mean degrees, respectively. The expected average degree is $c = (c_{\rm in} + c_{\rm out})/2$. The strength of the community structure is controlled by the parameter $\alpha = (c - c_{\rm out})/\sqrt{c}$: for $\alpha \leq 1$ the SBM is statistically indistinguishable from a graph without communities, while for larger values communities can be recovered by efficient algorithms \cite{decelle2011inference, krzakala2013spectral, dall2021unified}. In our experiments, we let $\alpha$ span from $0$ (which corresponds to an Erd\"os Rényi random graph) to $0.8\sqrt{c}$, ensuring a defined community structure.
    \item \textbf{Configuration Model (CM)} \cite{bollobas1980probabilistic}. This model generates random graphs with an arbitrary degree distribution. We construct the degree sequence via a logarithmic interpolation \cite{deserno2004linear} between a Gaussian distribution, $P_G(k)$, and a truncated power-law distribution, $P_{TPL}(k)$. The resulting probability mass function $P(k)$ for the degree sequence is defined as:
    \begin{align*}
    	P(k) \propto P_G(k)^{1-p_d} P_{TPL}(k)^{p_d}\,\,,
    \end{align*}
    where the parameter $p_d \in [0, 1]$ controls the heterogeneity of the network. Setting $p_d = 0$ yields a purely Gaussian degree distribution, resulting in a homogeneous network. Conversely, $p_d = 1$ produces a highly heterogeneous contact structure.
    \item \textbf{Watts-Strogatz (WS)} \cite{watts1998}. This model generates networks with simultaneously high clustering coefficients and short average path lengths, two properties commonly observed in real-world contact networks. The rewiring probability $p_r$ interpolates between a regular lattice (high clustering, large diameter) and a random graph (low clustering, small diameter). We vary $p_r$ in the range [0.01, 1].
\end{itemize}
For each graph type, we vary the controlling parameter ($\alpha$, $p_d$, or $p_r$) across 5 values, randomly select the number of nodes in the range $[300, 1500]$, and fix the expected average degree at 10.
We model contact dynamics by assigning temporal activation sequences to each edge via a probabilistic mixture controlled by $q \in [0, 1]$. With probability $q$, an edge receives an activation sequence sampled from the empirical SocioPatterns dataset \cite{school_dataset}, collected in a scholastic context; with probability $1-q$, its activation times are drawn from a Poisson process. Varying $q$ thus interpolates between purely synthetic Poissonian dynamics ($q=0$) and realistic bursty contact patterns ($q=1$).

For every unique combination of the 3 graph models, the 5 values of the controlling parameter, the 5 network sizes, and the 5 values of the temporal parameter $q$, we generate 10 independent network realizations. This systematic sampling strategy yields a total dataset of 3750 distinct synthetic temporal graphs.

\subsubsection{Results}
Following the same pipeline established for the empirical graphs, and applying our optimization procedure, we simulate the SIR epidemic spreading on these synthetic networks to compute the \emph{Time-Shift DTW} ground truth. The \acronym parameters $\bm{\gamma}$ are optimized exclusively on the training set, and the generalization performance is subsequently evaluated on the unseen test graphs. 

\begin{figure}[t!]
    \centering
    \includegraphics[width=\linewidth]{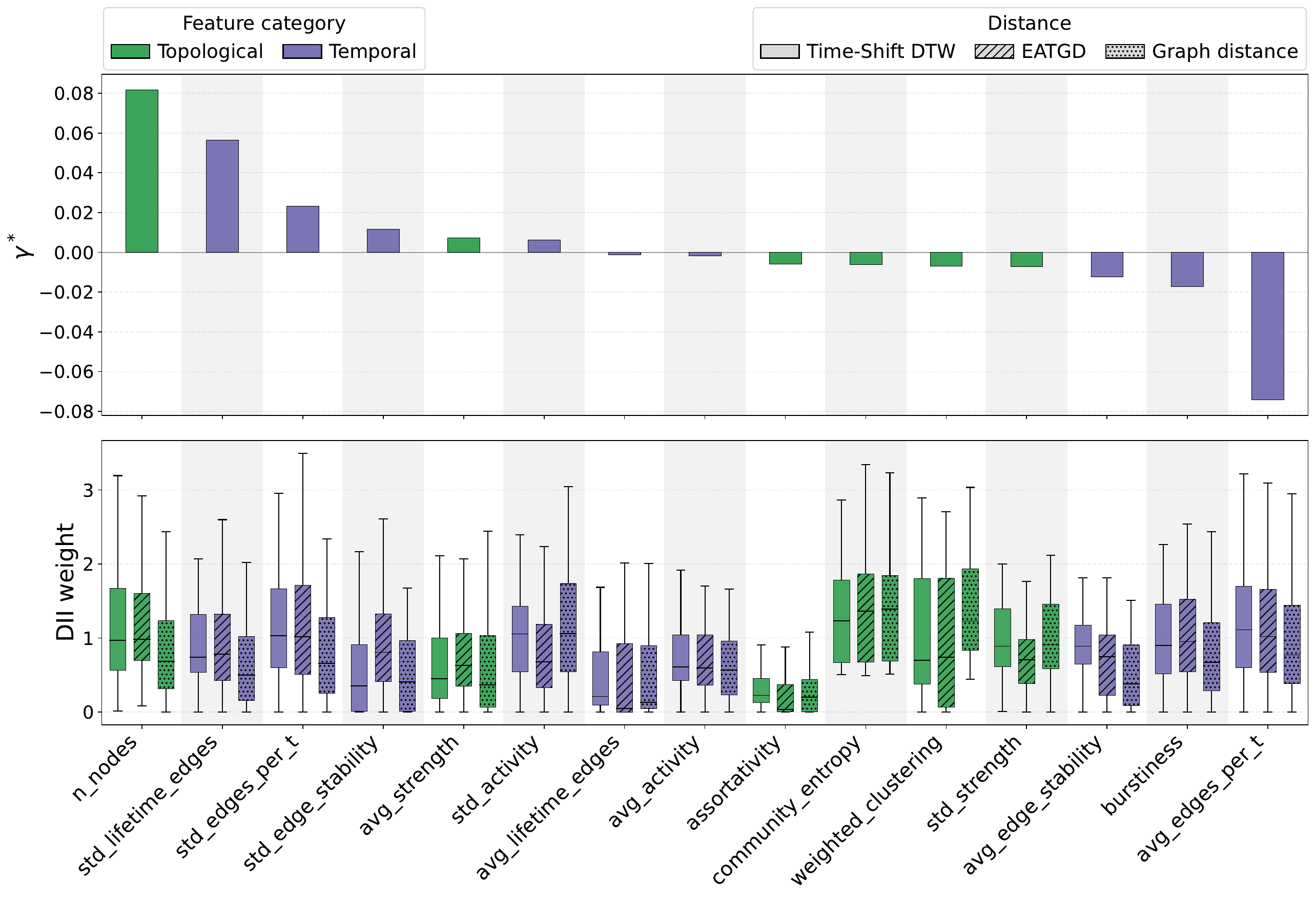}
    \caption{\textbf{Feature importance distributions across target measures (synthetic dataset).} (Top) Learned weights $\gamma_i^*$ for each graph feature obtained when calibrating EATGD on the synthetic training set, colored by feature category (topological vs. temporal). (Bottom) Comparison of the importance weights attributed by DII to each feature for Time-Shift DTW, the baseline temporal graph distance, and the optimized EATGD. Boxplots are obtained from 50 independent DII runs.}
    \label{fig:dii_synthetic}
\end{figure}

Because empirical human proximity patterns exhibit significantly different topological and temporal feature distributions compared to synthetic network models, the resulting optimal coefficients $\boldsymbol{\gamma}^*$ (Fig.~\ref{fig:dii_synthetic}) diverge from those learned on the SocioPatterns dataset. This variation highlights the adaptive capacity of the optimization framework, which automatically shifts its penalization according to the underlying data.

Our optimized measure achieves a correlation of $\rho = 0.54$ on the test set, starting from a baseline correlation of $\rho = 0.17$, confirming that EATGD scales to large datasets and successfully ranks synthetic temporal graphs by their similarity in epidemic spreading behavior.

As observed in the main text, the feature importance analysis through the DII shows consistency across the three measures (Fig. \ref{fig:dii_synthetic}).
This is evidenced by the total absolute Cohen's $d$, which drops from an initial baseline value of 2.34 (against the ground-truth \emph{Time-Shift DTW}) to 2.20, maintaining consistently low effect sizes across all features.

\subsection{Extended Robustness analysis}

In this section, we provide the extended results for the out-of-distribution (OOD) test discussed in the main text. Specifically, here we leave out pairs of datasets (environments) to serve as the test set, while training the learnable parameters on the remaining graphs. 

\begin{figure}[!h]
    \centering
    \includegraphics[width=0.74\textwidth]{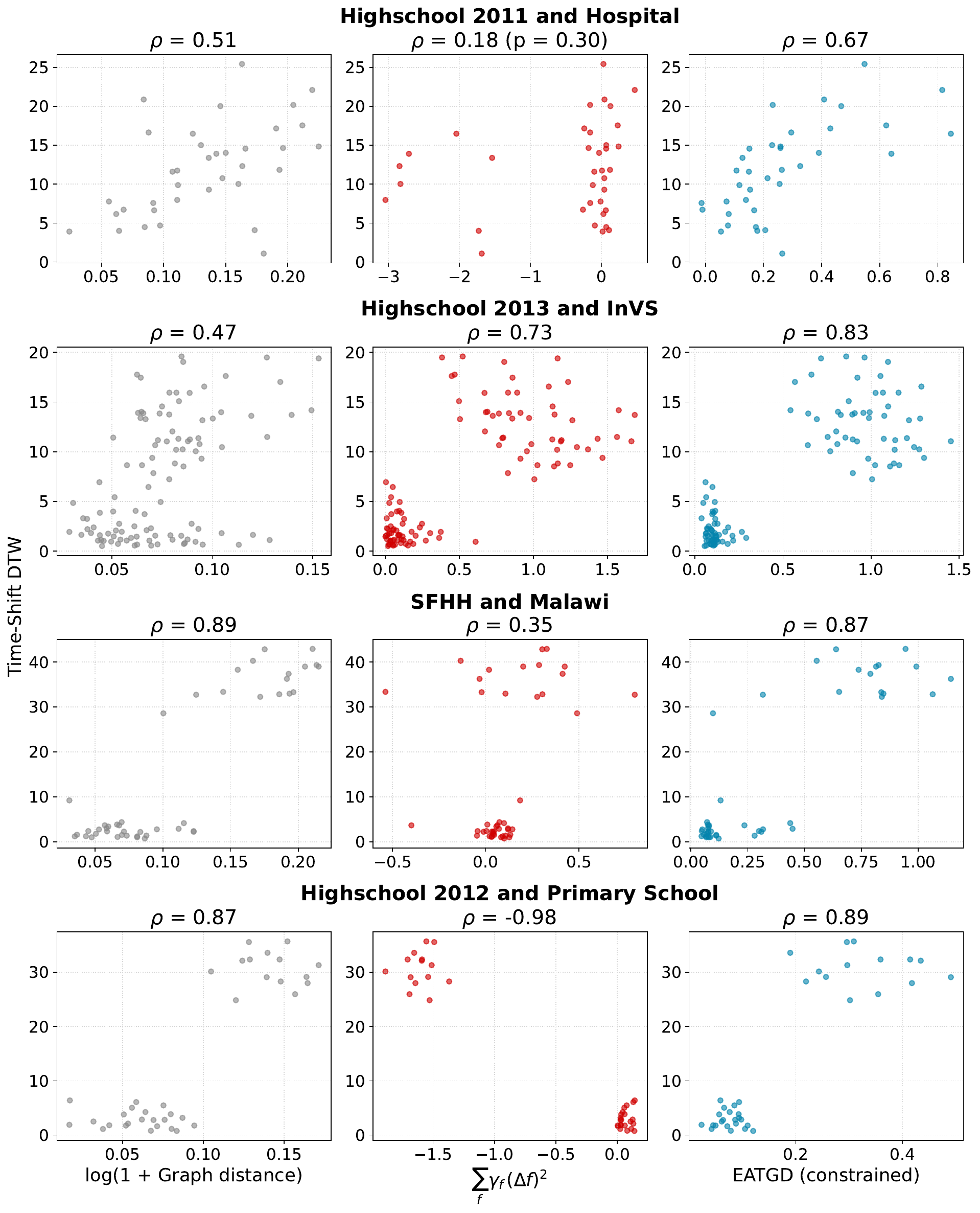}
    \caption{\textbf{Extended performance of EATGD in out-of-distribution scenarios.} Each row reports a different cross-environment evaluation: all temporal graphs from the two environments named above the row (\emph{e.g.}, \emph{Highschool 2011} and \emph{Hospital}) are held out as the test set, while EATGD is calibrated on the remaining SocioPatterns environments. The scatter plots compare the \emph{Time-Shift DTW} against: (left) the baseline topological distance, $\log[d(\mathcal{G}_1,\mathcal{G}_2)+1]$; (center) the unconstrained feature-based component of EATGD, $\sum_f \gamma_f(\Delta f)^2$, excluding the graph-distance term; and (right) the constrained EATGD, obtained by applying the statistical gating described in the main text. The unconstrained feature-based component is unstable out of distribution — in the most extreme case (\emph{Highschool 2012} and \emph{Primary School}, bottom row) it is anti-correlated with the \emph{Time-Shift DTW} ($\rho = -0.98$) — whereas the constrained EATGD consistently matches or outperforms the topological baseline across all four held-out pairs.}
    \label{fig:sm_gating_full}
\end{figure}

Figure~\ref{fig:sm_gating_full} reports the scatter plots comparing the \emph{Time-Shift DTW} against different configurations of our measure for different test sets. As anticipated in the main text, applying the unconstrained measure to unfamiliar domains can lead to severe numerical instabilities. For instance, in extreme OOD scenarios (e.g., \emph{Highschool 2012} + \emph{Primary School}), the unconstrained feature differences completely disrupt the structural alignment, leading to negative or non-significant correlations. Conversely, the application of a threshold on the absolute standardized feature values filters out these topological anomalies. This mechanism stabilizes the measure across the evaluated cases, allowing it to systematically match or outperform the purely topological baseline ($\log \left[d(\mathcal{G}_1, \mathcal{G}_2) +1 \right]$).

\end{document}